 \documentclass[final,3p,times,twocolumn]{elsarticle}

 \usepackage{graphics}
 \usepackage{graphicx}

\usepackage{amssymb}
\usepackage{amsmath}

\journal{Journal of Molecular Spectroscopy}

\begin{document}

\begin{frontmatter}



\title{Rotational spectroscopy of isotopic oxirane, $c$-C$_2$H$_4$O}


\author[Koeln]{Holger S.P.~M\"uller\corref{cor}}
\ead{hspm@ph1.uni-koeln.de}
\cortext[cor]{Corresponding author.}
\author[Rennes]{Jean-Claude Guillemin} 
\author[Koeln]{Frank Lewen} 
\author[Koeln]{Stephan Schlemmer} 

\address[Koeln]{Astrophysik/I.~Physikalisches Institut, Universit{\"a}t zu K{\"o}ln, 
  Z{\"u}lpicher Str. 77, 50937 K{\"o}ln, Germany}
\address[Rennes]{Univ Rennes, Ecole Nationale Sup{\'e}rieure de Chimie de Rennes, 
  CNRS, ISCR$-$UMR 6226, 35000 Rennes, France}

\begin{abstract}

We studied the rotational spectrum of oxirane in a sample of natural isotopic composition 
in selected regions between 158~GHz and 1093~GHz. Investigations of the isotopologs with 
one $^{13}$C or one $^{18}$O were the primary focus in order to facilitate searches for 
them in space. We also examined the main isotopic species mainly to look into the performance 
of Watson's A and S reductions both in an oblate and in a prolate representation. 
Even though oxirane is a rather asymmetric oblate rotor, the A reduction in the III$^l$ 
representation did not yield a satisfactory fit, as was observed frequently earlier for 
other molecules. The other three combinations yielded satisfactory fits of similar quality 
among each other; the A reduction in the I$^r$ representation required two parameters less 
than both S reduction fits.

\end{abstract}

\begin{keyword}  

rotational spectroscopy \sep 
submillimeter spectroscopy \sep 
interstellar molecule \sep
cyclic molecule \sep
centrifugal distortion \sep 
reduced Hamiltonian


\end{keyword}

\end{frontmatter}




\section{Introduction}
\label{introduction}

Oxirane, $c$-C$_2$H$_4$O, also know as ethylene oxide, oxacyclopropane, epoxyethane, 
and dimethylene oxide, is a molecule of astrochemical interest that was detected first 
toward the prolific high-mass Galactic center source Sagittarius (Sgr) B2(N) 
\cite{det-c-C2H4O_1997}. It was found subsequently in several other high-mass 
star-forming regions \cite{more_obs_c-C2H4O_1998,still_more_obs_c-C2H4O_2001}. 
It was also observed in three rotationally cold, but kinetically warm, Galactic center 
sources \cite{with-O_Miguel_2008} and more recently toward the prototypical low-mass 
protostellar source IRAS 16293$-$2422 \cite{propanal_etc_2017} and toward a prestellar 
core \cite{c-C2H4O_L1689B_2019}.

Observations of molecules containing $^{13}$C are viewed as important diagnostic tools 
because the $^{12}$C/$^{13}$C ratio in the interstellar medium (ISM) differs from the 
terrestrial value of 89 \cite{iso-comp_2011}. It is around 20 to 25 in the Galactic 
center region \cite{12C-13C_SgrB2N_2017,13C-VyCN_2008,EMoCA_2016,RSH_ROH_2016,det-13CH_2020}, 
increases to about 68 in the Solar neighborhood and even further in the outskirts of the Milky Way 
\cite{12C-13C_SgrB2N_2017,det-13CH_2020,galactic_isotopic-ratios_1994,12C-13C-gradient_2005}. 
Even lower $^{12}$C/$^{13}$C ratios were found in the envelopes of some late-type stars, 
such as K4$-$47 \cite{PN_K4-47_2019,PN_K4-47_12C-13C_etc_2018}.

Numerous detections of isotopologs containing $^{13}$C were reported in recent years 
\cite{12C-13C_SgrB2N_2017,RSH_ROH_2016,2x13C-EtCN_rot_det_2016,PILS_div-isos_2018,det-13CH_2020,isos-HCCNC_2020} 
and some even much earlier. Variations in the $^{12}$C/$^{13}$C ratios within one source, 
such as in IRAS 16293$-$2422 \cite{PILS_div-isos_2018}, may provide clues on the formation 
pathways of these molecules. Some of these observations benefited greatly from recent 
or concomitant laboratory investigations of $^{13}$C containing isotopologs, such as 
ethanol \cite{13C-EtOH_rot_2012}, acetaldehyde \cite{13C-MeCHO_rot_2015}, or ethyl 
cyanide with two $^{13}$C \cite{2x13C-EtCN_rot_det_2016}. Other recently studied 
isotopologs await detection in space. These include methylamine \cite{13CH3NH2_rot_2016}, 
methyl isocyanate \cite{CH3NCO_rot_2019}, methyl mercaptan \cite{13CH3SH_rot_2020}, 
and cyclopropenone \cite{c-H2C3O_rot_2021}.

The relative abundance of $^{18}$O is less favorable; the terrestrial $^{16}$O/$^{18}$O 
ratio is almost exactly 500 \cite{iso-comp_2011}, and the ratio in the Galactic center 
of $\sim$200 \cite{RSH_ROH_2016,galactic_isotopic-ratios_1994,isotopic_formaldehyde_1985,18O-MeOH_det_1989} 
is only slightly lower. Nevertheless, some complex organic molecules containing $^{18}$O 
have been detected in recent years, among them methyl formate \cite{18O-MeFo_det_2012} 
and formamide \cite{urea_ReMoCA_2019}. Cyclopropenone is an example of a molecule 
detected in the ISM whose $^{18}$O isotopolog was studied recently, but was not yet 
detected \cite{c-H2C3O_rot_2021}.

The first studies of the rotational spectrum of oxirane date back to the early days 
of microwave spectroscopy and include the determination of structural parameters and 
of its dipole moment through Stark effect measurements \cite{c-C2H4O_S_rot_dip_1951}. 
Its dipole moment was determined even earlier through investigations of its dielectric 
properties along with those of several other molecules \cite{c-C2H4O_etc_dip_1928}.
More accurate transition frequencies of several isotopic species were determined later 
in the microwave and lower millimeter-wave regions 
\cite{c-C2H4O_div-isos_rot_1974,c-C2H4O_rot_1974,c-C2H4O_rot_isos_1974}. 
Extensive data pertaining to the main isotopolog were obtained more recently 
between 260 and 360~GHz \cite{c-C2H4O_FASSST_1998} and between 15 and 73~cm$^{-1}$ 
\cite{c-C2H4O_FIR_2012}. Results of a millimeter-wave and far-infrared study on 
$c$-C$_2$H$_3$DO were reported very recently \cite{c-C2H3DO_rot_2019}. 
Unlabeled atoms in formulae designate $^{12}$C and $^{16}$O here and in the following.
We also mention a high-resolution IR study \cite{c-C2H4O_IR_2012} and intensity 
measurements on $c$-C$_2$H$_4$O \cite{c-C2H4O_IR_2014}.

Our present work describes the investigations of \mbox{$c$-}$^{13}$CCH$_4$O and 
$c$-C$_2$H$_4$$^{18}$O employing samples of oxirane in natural isotopic composition 
to facilitate their detection in space. We also measured transition frequencies of 
the main isotopic species mainly to test the performances of Watson's S and A 
reductions of the rotational Hamiltonian in combination with an oblate and a prolate 
representation.


\section{Experimental details}
\label{exptl_details}

\begin{figure}
\centering
\includegraphics[width=7cm,angle=0]{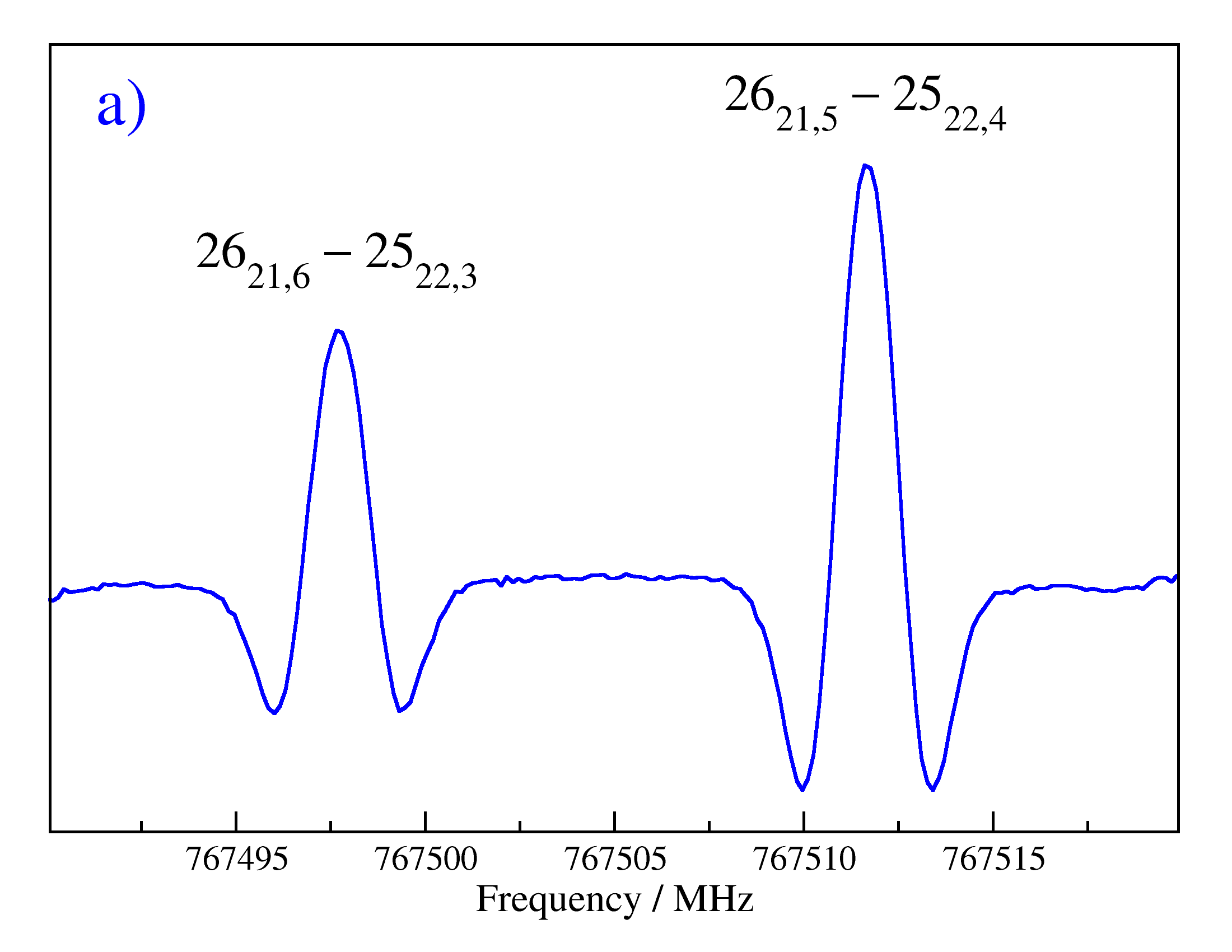}
\includegraphics[width=7cm,angle=0]{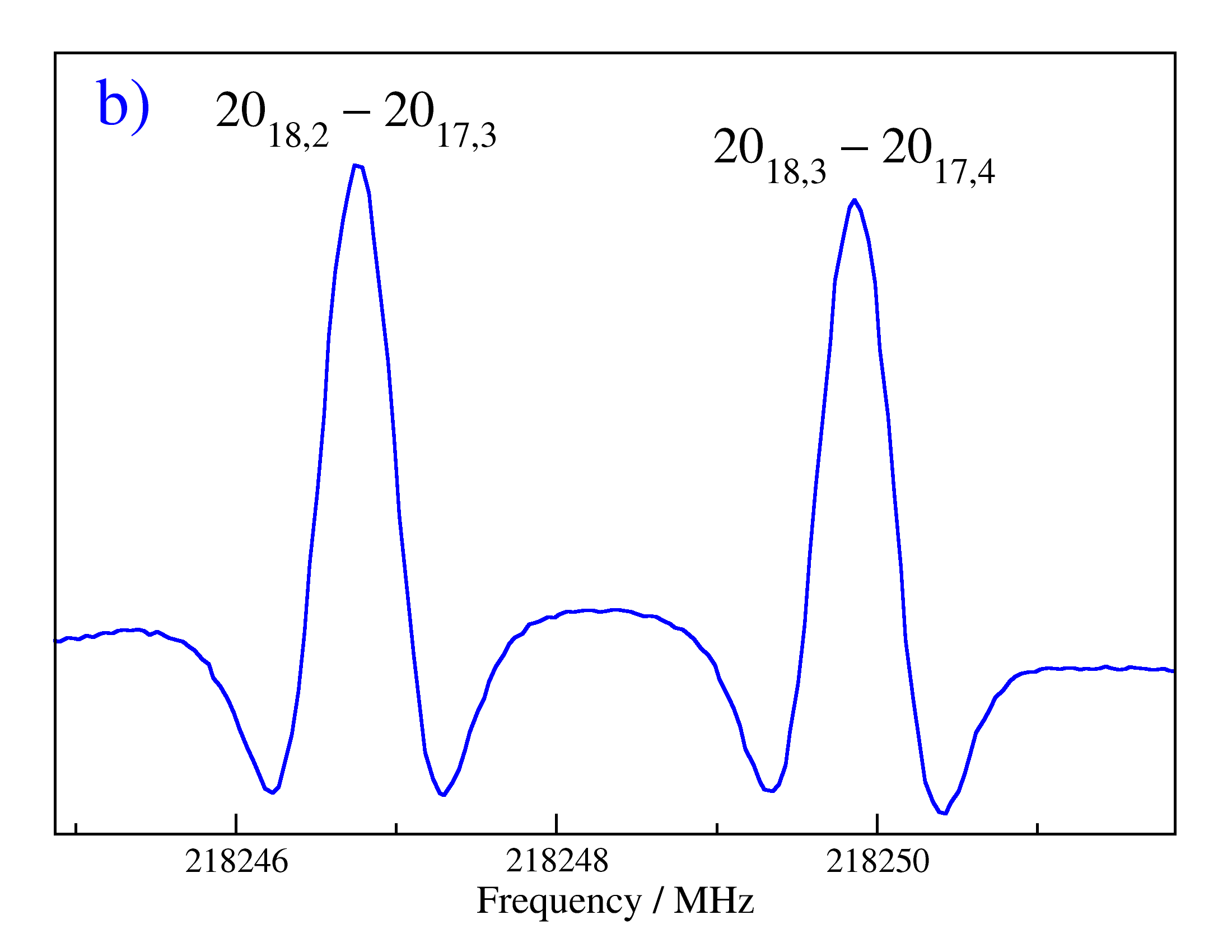}
\includegraphics[width=7cm,angle=0]{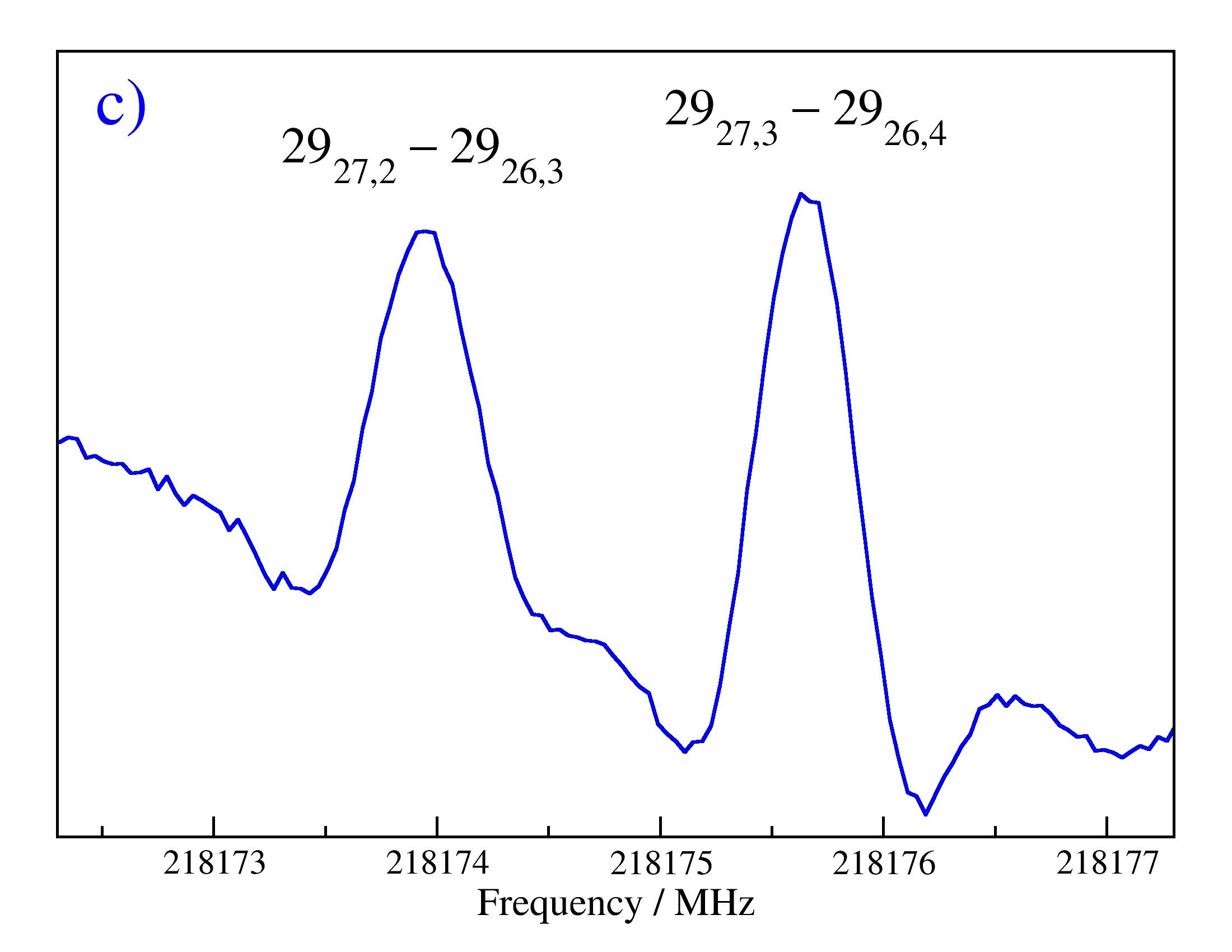}

\caption{Sections of the rotational spectrum of oxirane displaying the presence of 
   $3 : 5$ \textit{para} to \textit{ortho} spin-statistics for $c$-C$_2$H$_4$O (a) 
   and $c$-C$_2$H$_4$$^{18}$O (c) as well as the absence in $c$-$^{13}$CCH$_4$O (b).}
\label{spin-statistics}
\end{figure}

Our measurements were carried out at room temperature using two different spectrometers. 
Pyrex glass cells with an inner diameter of 100~mm were employed. Both spectrometer 
systems used VDI frequency multipliers driven by Rohde \& Schwarz SMF~100A microwave 
synthesizers as sources. Schottky diode detectors were utilized below 250~GHz, whereas 
a closed cycle liquid He-cooled InSb bolometer (QMC Instruments Ltd) was applied 
above 340~GHz. Frequency modulation was used throughout. The demodulation at $2f$ causes 
an isolated line to appear close to a second derivative of a Gaussian, as can be seen 
in Fig.~\ref{spin-statistics}.


A 5~m long double pass cell equipped with Teflon lenses was used for measurements between 
158 and 250~GHz. The pressures were about 1~Pa. Further information on this spectrometer 
is available elsewhere \cite{OSSO_rot_2015}. We achieved frequency accuracies of 5~kHz 
for this spectrometer in a study of 2-cyanobutane \cite{2-CAB_rot_2017}, a molecule that 
displays a much richer rotational spectrum. Measurements between 340 and 505~GHz and 
between 758 and 1093~GHz were carried out employing a 5~m long single pass cell at 
pressures of $\sim$2~Pa. A pressure slightly below 1~Pa was chosen for measurements 
of medium strong lines of the main isotopolog. 
Our studies on isotopic formaldehyde \cite{H2CO_rot_2017} or thioformaldehyde 
\cite{H2CS_rot_2019} demonstrate that accuracies of 10~kHz can be reached quite easily 
for very symmetric lines with good signal-to-noise ratio up to 1.5~THz. 
We assigned uncertainties of mostly 5 or 10~kHz for the best lines up to 50~kHz for 
less symmetric lines or lines fairly close to other lines.


\section{Results}
\label{results}

\subsection{Spectroscopic properties of oxirane}
\label{obs_spectrum}

Oxirane is a very asymmetric rotor of the oblate type with $\kappa = (2B - A - C)/(A - C)$ 
equals 0.4093, 0.3557, and 0.6389 for $c$-C$_2$H$_4$O, $c$-$^{13}$CCH$_4$O, and 
$c$-C$_2$H$_4$$^{18}$O, respectively. The main isotopolog as well as the one with $^{18}$O 
have $C_{\rm 2v}$ symmetry, and the four equivalent H atoms lead to \textit{para} and 
\textit{ortho} spin-statistics with a $3 : 5$ intensity ratio, see Fig.~\ref{spin-statistics}. 
The \textit{para} and \textit{ortho} levels are described by $K_a + K_c $ being odd and even, 
respectively. These spin-statistics are absent in $c$-$^{13}$CCH$_4$O because the molecule 
has only $C_{\rm S}$ symmetry.

The dipole moment of 1.90~D of $c$-C$_2$H$_4$O is along the $b$-axis \cite{c-C2H4O_S_rot_dip_1951}. 
This value includes a small increase of 0.02~D caused by the difference of the OCS reference value 
used in \cite{c-C2H4O_S_rot_dip_1951} compared to more modern values \cite{OCS_dip_1985,OCS_dip_1986}. 
The lower symmetry of $c$-$^{13}$CCH$_4$O is associated with a small $a$-dipole moment component 
of $\sim$0.17~D.

The rotational spectrum of oxirane is sparse at the level of the strongest lines because 
the rotational parameters are fairly large. However, the spectrum is much richer at 
the level of the weakest transitions in the line list which are about 4.5 orders of 
magnitude weaker, leading to somewhat increased accidental overlap of lines.

\subsection{Observed spectrum and determination of spectroscopic parameters of $c$-C$_2$H$_4$O}
\label{parameters-main}

The initial calculation of the rotational spectrum of $c$-C$_2$H$_4$O was based on 
the second version of the Cologne Database for Molecular spectroscopy, CDMS, 
\cite{CDMS_2001,CDMS_2005,CDMS_2016} entry of the oxirane main isotopic species
from July 2012, which, in turn, is based on the FIR study from that year 
\cite{c-C2H4O_FIR_2012}. This study employed Watson's A reduced Hamiltonian 
in the prolate I$^r$ representation, as was done in the first version of the 
$c$-C$_2$H$_4$O CDMS catalog entry of March 2000, on which the FIR study built. 
The same combination of reduction and representation was used in an IR study 
\cite{c-C2H4O_IR_2012}, whereas Pan et al. employed Watson's A reduction in the 
oblate III$^l$ representation using 25 spectroscopic parameters in their fit 
\cite{c-C2H4O_FASSST_1998}. The initial CDMS fit, in contrast, used only 14 
spectroscopic parameters, and the FIR study as well as the second CDMS fit 
used only one additional spectroscopic parameter.


\begin{table*}
\begin{center}
\caption{Spectroscopic parameters$^a$ (MHz) of the main isotopolog of oxirane employing Watson's S and A reduction of the rotational 
         Hamiltonian both in the oblate representation III$^l$ and in the prolate representation I$^r$.}
\label{tab-main-species}
\renewcommand{\arraystretch}{1.08}
\begin{tabular}[t]{lr@{}lr@{}lcr@{}lr@{}ll}
\hline \hline
 \multicolumn{5}{c}{S reduction} & & \multicolumn{5}{c}{A reduction} \\
\cline{1-5} \cline{7-11}
Parameter & \multicolumn{2}{c}{III$^l$} & \multicolumn{2}{c}{I$^r$} & & 
\multicolumn{2}{c}{I$^r$} & \multicolumn{2}{c}{III$^l$} & Parameter \\
\hline
$A$                      &  25483&.88987~(11)  &   25483&.88751~(10) & & 25483&.86273~(9)   &   25483&.88386~(10) & $A$                       \\
$B$                      &  22120&.83638~(9)   &   22120&.82714~(9)  & & 22120&.87303~(7)   &   22120&.84394~(9)  & $B$                       \\
$C$                      &  14097&.83590~(11)  &   14097&.85210~(11) & & 14097&.82581~(8)   &   14097&.83400~(16) & $C$                       \\
$D_K \times 10^3$        &     26&.07490~(13)  &       2&.95932~(23) & &    27&.58930~(22)  &      27&.63421~(33) & $\Delta_K \times 10^3$    \\
$D_{JK} \times 10^3$     &  $-$68&.65174~(30)  &      50&.46597~(26) & &    20&.91041~(21)  &   $-$70&.50171~(68) & $\Delta_{JK} \times 10^3$ \\
$D_J \times 10^3$        &     50&.83976~(24)  &      15&.75678~(24) & &    20&.68240~(14)  &      51&.14486~(26) & $\Delta_J \times 10^6$    \\
$d_1 \times 10^3$        &      9&.02067~(5)   &    $-$6&.21001~(6)  & &    18&.10824~(9)   &       3&.39130~(43) & $\delta_K \times 10^3$    \\
$d_2 \times 10^3$        &   $-$0&.15233~(2)   &    $-$2&.46284~(1)  & &     6&.21005~(2)   &    $-$9&.01957~(5)  & $\delta_J \times 10^3$    \\
$H_K \times 10^9$        & $-$191&.41~(11)     &    2712&.32~(23)    & &     2&.2415~(2)    &   $-$10&.8512~(10)  & $\Phi_K \times 10^6$      \\
$H_{KJ} \times 10^9$     &    373&.61~(21)     & $-$3535&.95~(28)    & &  $-$2&.6387~(3)    &      16&.0605~(16)  & $\Phi_{KJ} \times 10^6$   \\
$H_{JK} \times 10^9$     & $-$198&.52~(24)     &    1096&.22~(23)    & &   624&.58~(18)     & $-$5321&.27~(97)    & $\Phi_{JK} \times 10^9$   \\
$H_J \times 10^{9}$      &     17&.92~(19)     &   $-$46&.58~(18)    & &  $-$1&.797~(69)    &     110&.91~(22)    & $\Phi_J \times 10^{9}$    \\
$h_1 \times 10^{9}$      &   $-$8&.211~(44)    &       5&.268~(45)   & &  $-$0&.3278~(1)    &   $-$18&.0197~(18)  & $\phi_K \times 10^6$      \\
$h_2 \times 10^{9}$      &     46&.364~(24)    &      22&.470~(14)   & &   277&.30~(10)     &    1546&.63~(67)    & $\phi_{JK} \times 10^9$   \\
$h_3 \times 10^{9}$      &  $-$49&.529~(4)     &    $-$6&.868~(3)    & &  $-$1&.730~(7)     &   $-$57&.773~(50)   & $\phi_J \times 10^{9}$    \\
$L_K \times 10^{12}$     &   $-$3&.325~(25)    &   $-$14&.450~(79)   & &  $-$0&.62~(9)      &  $-$132&.5~(38)     & $L_K \times 10^{12}$      \\
$L_{KKJ} \times 10^{12}$ &      8&.202~(86)    &      26&.454~(117)  & &  $-$6&.79~(14)     &     177&.5~(61)     & $L_{KKJ} \times 10^{12}$  \\
$L_{JK} \times 10^{12}$  &   $-$8&.000~(98)    &   $-$13&.601~(102)  & &     8&.18~(11)     &   $-$33&.5~(23)     & $L_{JK} \times 10^{12}$   \\
$L_{JJK} \times 10^{12}$ &      4&.200~(57)    &    $-$1&.218~(58)   & &  $-$3&.00~(5)      &    $-$9&.36~(40)    & $L_{JJK} \times 10^{12}$  \\
$L_J \times 10^{12}$     &   $-$1&.083~(49)    &       0&.469~(44)   & &      &             &    $-$1&.31~(6)     & $L_J \times 10^{12}$      \\
$l_1 \times 10^{15}$     &    479&.6~(119)     &     117&.3~(99)     & & $-$13&.00~(4)      &  $-$207&.8~(48)     & $l_K \times 10^{12}$      \\
$l_2 \times 10^{15}$     &  $-$72&.7~(81)      &  $-$297&.1~(36)     & &     8&.07~(5)      &  $-$110&.5~(14)     & $l_{KJ} \times 10^{12}$   \\
$l_3 \times 10^{15}$     &     87&.5~(25)      &  $-$147&.8~(15)     & &  $-$2&.42~(3)      &       4&.43~(26)    & $l_{JK} \times 10^{12}$   \\
$l_4 \times 10^{15}$     &  $-$11&.9~(4)       &      31&.7~(2)      & &      &             &       0&.561~(14)   & $l_J \times 10^{12}$      \\
                         &       &             &        &            & &      &             &  $-$187&.5~(9)      & $P_K \times 10^{15}$      \\
                         &       &             &        &            & &      &             &     361&.2~(30)     & $P_{KKJ} \times 10^{15}$  \\
                         &       &             &        &            & &      &             &  $-$204&.5~(33)     & $P_{KJ} \times 10^{15}$   \\
                         &       &             &        &            & &      &             &      30&.7~(10)     & $P_{JK} \times 10^{15}$   \\
                         &       &             &        &            & &      &             &   $-$33&.9~(9)      & $p_K \times 10^{15}$      \\
                         &       &             &        &            & &      &             &      90&.7~(23)     & $p_{KKJ} \times 10^{15}$  \\
                         &       &             &        &            & &      &             &   $-$12&.7~(3)      & $p_{JK} \times 10^{15}$   \\
\# trans.$^{b,c}$        &   2942&             &    2942&            & &  2942&             &    2942&            & \# trans.$^{b,c}$         \\
\# new tr.$^{b,c}$       &   1091&             &    1091&            & &  1091&             &    1091&            & \# new tr.$^{b,c}$        \\
$J_{\rm max}$$^{b,c}$    &     63&             &      63&            & &    63&             &      63&            & $J_{\rm max}$$^{b,c}$     \\
$K_{a,\rm max}$$^{b,c}$  &     43&             &      43&            & &    43&             &      43&            & $K_{a,\rm max}$$^{b,c}$   \\
$K_{c,\rm max}$$^{b,c}$  &     61&             &      61&            & &    61&             &      61&            & $K_{c,\rm max}$$^{b,c}$   \\
wrms$^{b,c}$             &      0&.940         &       0&.957        & &     0&.946         &       1&.162        & wrms$^{b,d}$              \\
\hline
\end{tabular}\\[2pt]
\end{center} 
$^a$Numbers in parentheses are one standard deviation in units of the least significant figures.\\
$^b$Dimensionless.\\
$^c$See Section~\ref{parameters-main} for additional details on the numbers of transitions and the maximum quantum numbers.\\
$^d$Weighted standard deviation of the fit.
\end{table*}


We scrutinized the existing line lists in order to approach the best possible 
spectroscopic parameters as closely as possible. The FIR transition frequencies 
had uncertainties of 0.00020~cm$^{-1}$ assigned initially \cite{c-C2H4O_FIR_2012}. 
These uncertainties were judged rather conservatively given their rms value was 
0.00011~cm$^{-1}$ in the fit of the FIR study \cite{c-C2H4O_FIR_2012}. 
In addition, one pair of transitions, $J = 64 - 63$, $K_c = 62 - 61$ at 
61.468827~cm$^{-1}$, was omitted from our fit because of a large residual of 
$\sim$0.0012~cm$^{-1}$ between measured and calculated frequencies, and one pair 
of transitions, $J = 61 - 60$, $K_c = 61 - 60$ supposedly at 57.435516~cm$^{-1}$, 
was reassigned to a much stronger transition ($37_{26,12} - 36_{25,11}$) much closer 
to the calculated position. The FIR rms was reduced to 0.00009~cm$^{-1}$ after 
these two corrections, and we assigned 0.00010~cm$^{-1}$ as uncertainties to the 
FIR transition frequencies. The assignments of four more transition frequencies were 
extended eventually because more individual transitions contributed substantially to 
each measured line than was indicated in the initial line list \cite{c-C2H4O_FIR_2012}; 
further details are available in our line list which is part of the Supplementary Material. 
These corrections, however, had only very small effects on the fit. It is important 
in this context to emphasize that in the case of two or more overlapping transitions, 
our fitting program SPFIT only judges the intensity-weighted average position of all 
overlapping transitions.

One set of MW transition frequencies \cite{c-C2H4O_rot_isos_1974} was reportedly accurate 
to 0.01~MHz, but several lines displayed larger residuals between calculated and measured 
frequencies. In fact, the rms of this data set was very close to 0.02~MHz, and we attributed 
this value to this set of transition frequencies. Uncertainties of 0.05~MHz and 0.15~MHz, 
respectively, were employed for the remaining datasets \cite{c-C2H4O_rot_1974,c-C2H4O_FASSST_1998}, 
as was done earlier \cite{c-C2H4O_FIR_2012}.

Spectroscopic parameters were determined for this as well as various intermediate line lists 
using the A reduction in the I$^r$ representation and employing the S reduction in the III$^l$ 
and I$^r$ representations. The A reduction in the III$^l$ representation was only used for the 
final line list as this combination of reduction and representation appeared to require many 
more spectroscopic parameters than the other combinations. 
Pickett's SPFIT program \cite{spfit_1991} was applied for all fitting, and the SPCAT program 
was utilized for calculations of all rotational spectra.

We recorded individual transitions or pairs thereof if transitions happened to be close 
in frequency throughout our study. Series in particular combinations of $K_a$ and $K_c$ 
were followed in all instances. Asymmetry doublets with significant, but not complete 
splitting were avoided as their positions are usually determinable with less accuracy 
compared to well collapsed or well resolved asymmetry doublets. 
The strongest transitions of the main isotopolog were avoided because their uncertainties 
were already small such that their impact in the fit would be small. Moreover, these 
transitions are affected by opacity issues.

The 340$-$506~GHz region was studied extensively with 842 transitions in the final line list. 
Additional measurements were made between 160$-$250~GHz and 764$-$1046~GHz, where 61 and 188 
further transitions, respectively, were recorded and retained in the final line list. 
The number of different frequencies is 727; 728 transitions represent unresolved asymmetry 
doublets. The maximum quantum numbers $J$, $K_a$, and $K_c$ of our new transition frequencies 
are 55, 43, and 52, respectively.

The newly recorded transition frequencies deviated by modest amounts from the calculations 
based on the previous FIR study \cite{c-C2H4O_FIR_2012}, the largest deviation was 4.32~MHz. 
The deviations were largest for transitions with high values of $J$ and with $K_c = J$. 
The rms of the new lines fit with the old parameters was 580~kHz, which corresponded to 
38 times the experimental uncertainties on average. The signed deviation was 15~kHz, 
which means that deviations to higher frequencies were only marginally more important 
than deviations to lower frequencies.

We test the need for additional spectroscopic parameters usually by adding one parameter at 
a time and evaluating which of the new parameters improves the fit the most. This procedure 
ensures that the parameter set is about as small as possible and that it is fairly unique. 
The procedure works usually very well for prolate rotors, though correlation may in some cases 
require sets of two parameters to be tested. Fitting oblate type rotors was often less 
straightforward. In all combinations of reductions and representations we tested sets of 
two parameters to fit the $c$-C$_2$H$_4$O transition frequencies; we even tested sets of 
three and four parameters in the case of the A reduction in the III$^l$ representation.

The previous FIR study \cite{c-C2H4O_FIR_2012} employed 15 parameters using the A reduction 
in the I$^r$ representation; these were a nearly complete set of parameters up to sixth 
order, except $\Phi_J$, plus $l_K$. Our corresponding new fit consists of seven more parameters 
and is described as a nearly complete set of parameters up to eighth order, except $L_J$ and 
$l_J$, as summarized in Table~\ref{tab-main-species}. The weighted rms (or rms error) of the 
fit is 0.946 with modest scatter for the individual data sets: 1.087 and 1.057 for 19 
\cite{c-C2H4O_rot_isos_1974} and 40 microwave lines \cite{c-C2H4O_rot_1974}, respectively; 
1.000 for 607 millimeter and submillimeter transitions \cite{c-C2H4O_FASSST_1998}; 
0.872 for 1185 FIR transitions \cite{c-C2H4O_FIR_2012}; and 0.957 for our 1091 transitions 
with an rms of 20.3~kHz.

The quality of the fits utilizing the S reduction in the III$^l$ or the I$^r$ representation 
were similar, but required two parameters more in most fits; the difference was in some 
intermediate fits larger, in few smaller. The final spectroscopic parameters are also listed 
in Table~\ref{tab-main-species}. We tested the predictive power of the three parameter sets 
occasionally and found that there was no persistent preference for either set.

Obtaining a somewhat satisfactory fit applying the A reduction in the III$^l$ representation 
with the final line list was challenging. A complete set of parameters up to the eighth order 
plus several decic parameters yielded a fit that had an rms error more than 20\% worse than 
any of the other fits; the parameters of this fit are also given in Table~\ref{tab-main-species}. 
We tested all reasonable parameters individually, all or nearly all combinations of two of 
these along with several combinations of three and even some four parameter combinations. 
These improved the rms error only marginally, $\sim$1.12 was the best we were able to achieve. 
This modest refinement was discarded as it came at the expense of many more spectroscopic 
parameters.

\subsection{Observed spectrum and determination of spectroscopic parameters 
            of $c$-$^{13}$CCH$_4$O and $c$-C$_2$H$_4$$^{18}$O}
\label{parameters-isos}

Only 11 and 15 transitions were reported by Creswell and Schwendeman for $c$-$^{13}$CCH$_4$O and 
$c$-C$_2$H$_4$$^{18}$O, respectively. The uncertainties were reportedly 10~kHz, a value that was 
used in early fits, but later increased to 20~kHz because of the rms of these data, as in the 
case of the main isotopolog, see section~\ref{parameters-main}. 
Data reported by Hirose \cite{c-C2H4O_div-isos_rot_1974} were not considered initially even though 
slightly more transition frequencies were reported extending to slightly higher quantum numbers. 
The reasons were the larger uncertainties and in part large residuals between measured and calculated 
frequencies already in the publication, especially for $c$-C$_2$H$_4$$^{18}$O.

The number of transition frequencies was small for both isotopologs. Therefore, the spectroscopic 
parameters of the main isotopic species in the A reduction and I$^r$ representation from the previous 
FIR study \cite{c-C2H4O_FIR_2012} were taken as starting values for both isotopic species. 
The rotational parameters were adjusted in a first step. Through several trial fits, $\delta_K$ and 
$\delta_J$ were determined as the parameters which improved the rms error of the $c$-$^{13}$CCH$_4$O 
fit the most. The corresponding parameters for $c$-C$_2$H$_4$$^{18}$O were also $\delta_K$ and 
$\delta_J$ and additionally $\Delta_{JK}$. Minor adjustments of some further parameters, as 
frequently done \cite{c-H2C3O_rot_2021,H2CO_rot_2017,H2CS_rot_2019}, improved the fit only slightly.

We started our investigations of the two minor isotopic species in the 340$-$505~GHz region 
followed by 158$-$250~GHz and 758$-$1092~GHz, as for the main isotopic species. There was no caution 
required to avoid strong transitions of these isotopologs as their abundances in natural isotopic 
composition are lower than those of the main species by factors of $\sim$45 (there are two 
structural identical C atoms in the molecule) and $\sim$500, respectively. The quality of the 
extrapolations to submillimeter wavelengths was unclear because of the limited data set. 
Moreover, only one $R$-branch transition frequency was reported for either isotopolog, and this 
was the $J = 1 - 0$ transition. Since patterns of two transitions are easier to recognize than 
individual transitions, we searched for transitions with small, but resolved asymmetry splitting. 
In the case of $c$-$^{13}$CCH$_4$O, these were transitions with $K_a = J$ and $J'' = 9$, 8, and 7 
as well as transitions having $K_c = J - 4$ and $J'' = 14$, 13, and 12. Both types of transitions 
were also searched for in the case of $c$-C$_2$H$_4$$^{18}$O, but with slightly different values 
of $J''$ because of the very different $\kappa$ value, see section~\ref{obs_spectrum}, namely 
10 to 7 for the nearly prolate paired transitions and 11 to 9 for the nearly oblate paired 
transitions plus the oblated paired $J'' = 14$ in this series. After almost all of these 
transitions were found to be unblended, the spectroscopic parameters of both isotopologs 
were updated, and more transitions with similar combinations of quantum numbers were recorded. 
Subsequently, further series were sought, including large series of $Q$-branch transitions 
until there were no more transitions that had enough intensity to be recorded and a calculated 
uncertainty at least of order of the achievable measurement accuracy.


\begin{table}
\begin{center}
\caption{Spectroscopic parameters$^a$ (MHz) of the oxirane isotopologs containing 
         one $^{13}$C or one $^{18}$O.}
\label{tab-isos}
\renewcommand{\arraystretch}{1.10}
\begin{tabular}[t]{lr@{}lr@{}l}
\hline \hline
Parameter & \multicolumn{2}{c}{$c$-$^{13}$CCH$_4$O} & \multicolumn{2}{c}{$c$-C$_2$H$_4$$^{18}$O} \\
\hline
$A$                       & 25291&.99778~(17) & 23992&.43450~(20) \\
$B$                       & 21597&.97459~(13) & 22121&.17197~(14) \\
$C$                       & 13825&.77925~(14) & 13628&.24006~(9)  \\
$\Delta_K \times 10^3$    &    27&.62104~(47) &    24&.65010~(70) \\
$\Delta_{JK} \times 10^3$ &    20&.47198~(31) &    17&.45625~(69) \\
$\Delta_J \times 10^6$    &    20&.03716~(31) &    20&.48033~(8)  \\
$\delta_K \times 10^3$    &    18&.29406~(30) &    16&.34518~(30) \\
$\delta_J \times 10^3$    &     6&.05179~(5)  &     6&.31363~(4)  \\
$\Phi_K \times 10^6$      &     1&.9510~(4)   &     2&.0748~(8)   \\
$\Phi_{KJ} \times 10^6$   &  $-$2&.2614~(6)   &  $-$2&.4735~(9)   \\
$\Phi_{JK} \times 10^6$   &     0&.5141~(2)   &     0&.5999~(6)   \\
$\Phi_J \times 10^{9}$    &     4&.97~(30)    &  $-$1&.9          \\
$\phi_K \times 10^6$      &  $-$0&.2208~(6)   &  $-$0&.3603~(3)   \\
$\phi_{JK} \times 10^6$   &     0&.2292~(5)   &     0&.2704~(3)   \\
$\phi_J \times 10^{9}$    &  $-$0&.251~(29)   &  $-$1&.6          \\
$L_K \times 10^{12}$      &  $-$0&.51         &  $-$0&.56         \\
$L_{KKJ} \times 10^{12}$  &  $-$7&.42~(37)    &  $-$6&.56         \\
$L_{JK} \times 10^{12}$   &     7&.00         &     7&.40         \\
$L_{JJK} \times 10^{12}$  &  $-$2&.40         &  $-$2&.48         \\
$L_J \times 10^{12}$      &  $-$1&.158~(97)   &      &            \\
$l_K \times 10^{12}$      & $-$12&.66~(35)    & $-$10&.1          \\
$l_{KJ} \times 10^{12}$   &     9&.45~(45)    &     7&.0          \\
$l_{JK} \times 10^{12}$   &  $-$3&.12~(21)    &  $-$2&.5          \\
$l_J \times 10^{12}$      &      &            &      &            \\
\# trans.$^{b,c}$         &  1137&            &   701&            \\
\# new tr.$^{b,c}$        &  1111&            &   667&            \\
$J_{\rm max}$$^{b,c}$     &    45&            &    38&            \\
$K_{a,\rm max}$$^{b,c}$   &    34&            &    30&            \\
$K_{c,\rm max}$$^{b,c}$   &    39&            &    38&            \\
wrms$^{b,d}$              &     0&.907        &     0&.906        \\
\hline
\end{tabular}\\[2pt]
\end{center} 
$^a$Watson's A reduction was used in the I$^r$ representation. Numbers in parentheses are 
    one standard deviation in units of the least significant figures. Spectroscopic parameters 
    without uncertainties were evaluated from the main isotopic species and kept fixed in the 
    fits, see section~\ref{parameters-isos}.\\
$^b$Dimensionless.\\
$^c$See Section~\ref{parameters-isos} for additional details on the numbers of transitions and the maximum quantum numbers.\\
$^d$Weighted standard deviation of the fit.
\end{table}


The final line lists consisted of 553 and 274 transitions between 340$-$505~GHz, 211 and 186 
transitions between 158$-$250~GHz, and 341 and 207 transitions between 758$-$1092~GHz for the 
isotopologs containing one $^{13}$C and one $^{18}$O, respectively. 
680 of the 1111 and 468 of the 667 transitions correspond to unresolved asymmetry doublets 
for $c$-$^{13}$CCH$_4$O and $c$-C$_2$H$_4$$^{18}$O, respectively, resulting in 771 and 433 
different transition frequencies. The maximum quantum numbers $J$, $K_a$, and $K_c$ of our 
new transition frequencies are 45, 34, and 39 for $c$-$^{13}$CCH$_4$O and 38, 30, and 38 
for $c$-C$_2$H$_4$$^{18}$O.

As the A reduction in the I$^r$ representation required the least number of spectroscopic 
parameters to fit the transition frequencies of the main isotopic species, we only tried this 
combination of reduction and representation. Additional parameters for the main isotopolog 
were subsequently transferred as fixed parameters to the fits of the minor isotopologs. 
Higher order distortion parameters of these, which were kept fixed in the fit, were 
adjusted if lower order parameters differed markedly from those of the main isotopic species.
The changes were applied by evaluating trends in related parameters, such as $\Delta_K$, 
$\Phi_K$, and $L_K$.

Our new transition frequencies of the two minor isotopic species deviated much more from 
the initial calculations than in the case of the main isotopolog because of the much smaller 
initial line lists for both minor isotopologs which resulted in substantially less reliable 
spectroscopic parameters. The rms of the new lines fit with the initial parameters is 
29.9~MHz, the average shift is 17.3~MHz, and the rms corresponds to more than 2500 times 
the experimental uncertainties on average for $c$-$^{13}$CCH$_4$O. In the case of 
$c$-C$_2$H$_4$$^{18}$O, that rms is 63.4~MHz, the average shift is $-$14.5~MHz, and 
the rms corresponds to almost 3450 times the experimental uncertainties on average.

Spectroscopic parameters derived from the main isotopolog or new spectroscopic parameters 
were tested as was done for the main isotopic species. That is, we searched for the parameter 
whose floating or inclusion improved the fit the most in each fitting round. An additional 
prerequisite to keep a new parameter in the fit or to keep a previously fixed parameter 
floated was that the parameter was determined sufficiently well, meaning that its uncertainty 
should be less than one fifth of its magnitude.

In the final fitting round, we tested the transition frequencies reported by Hirose 
\cite{c-C2H4O_div-isos_rot_1974}; 15 and 18 lines were included with uncertainties of 
0.05~MHz for $c$-$^{13}$CCH$_4$O and $c$-C$_2$H$_4$$^{18}$O, respectively, and 
three respectively four lines were omitted because of large residuals. 
The final sets of spectroscopic parameters are given in Table~\ref{tab-isos}.

The weighted rms (or rms errors) of the fits are 0.907 and 0.906 for the isotopologs 
containing one $^{13}$C and one $^{18}$O, respectively. The rms errors for 11 lines 
from Creswell and Schwendemann \cite{c-C2H4O_rot_isos_1974}, 15 lines from Hirose 
\cite{c-C2H4O_div-isos_rot_1974}, and 1111 lines from our present study are 0.898, 
0.805, and 0.909, respectively for $c$-$^{13}$CCH$_4$O; the rms of our lines is 
22.1~kHz. The corresponding values for 16 lines from Creswell and Schwendemann 
\cite{c-C2H4O_rot_isos_1974}, 18 lines from Hirose \cite{c-C2H4O_div-isos_rot_1974}, 
and 667 lines from our present study are 1.056, 0.978, and 0.897, respectively; 
the rms of our lines is 24.4~kHz.


\section{Discussion}
\label{discussion}

The main interest for our study of the oxirane main isotopolog was the test of various 
combinations of reduction and representation. We point out that we did not have any program 
at our disposal to try out the II$^r$ or II$^l$ representation, but these are very rarely 
used in published fits anyhow. A rather large set of spectroscopic parameters was needed 
to obtain a somewhat satisfactory fit in the III$^l$ representation using the A reduction, 
similar to results from a millimeter and submillimeter study \cite{c-C2H4O_FASSST_1998}. 
This combination of reduction and representation would be considered by many spectroscopists 
to be the natural choice as they commonly use the A reduction for traditional reasons and 
as oxirane is an asymmetric top of the oblate type that is with $\kappa = 0.4093$ quite far 
from the symmetric limit.

Considerably fewer parameters, however, are required in the III$^l$ representation using the 
S reduction, as can be seen in Table~\ref{tab-main-species}, emphasizing again the greater 
versatility of the S reduction compared to the A reduction if the representation is used 
that fits to the asymmetry type of the molecule. We may, of course, also use a different 
representation. The S reduction in the I$^r$ representation yields a fit of similar quality 
and with the same number of parameters as the S reduction in the III$^l$ representation. 
It is interesting to note that the A reduction in the I$^r$ representation requires two 
parameters fewer still to produce a fit of about the same quality. Thus, while the 
S reduction is the prime choice if one only considers the III$^l$ representation, 
the A reduction in the I$^r$ representation is preferable if all these four combinations 
are considered.

The poor performance of the A reduction in an oblate representation (III$^l$ and III$^r$ 
differ only in some signs of off-diagonal distortion parameters) is not unique to oxirane, 
but is at least somewhat more widespread. Yamada and Klee \cite{H2S_FIR_1994} carried out 
an FIR study on H$_2$S and fit their data together with rotational data of microwave 
accuracy, employing full parameter sets of eighth order and diagonal decic parameters. 
Although H$_2$S is an asymmetric rotor of the oblate type ($\kappa = 0.5234$), they 
encountered convergence problems using the A reduction in the III$^r$ representation. 
A satisfactory fit resulted from a fit in which the S reduction in the same representation 
was applied. The authors also tried fits in the I$^r$ representation. In this case, the A 
reduction gave a fit that was noticeably worse than the S and III$^r$ combination, 
whereas the S reduction (in I$^r$) yielded the best fit.

Another example is the lowest energy conformer of 2-cyanobutane, which is a very asymmetric 
rotor of the oblate type $\kappa = 0.1404$ \cite{2-CAB_rot_2017}. The combination of 
A reduction and III$^l$ representation required the most parameters; the best result was 
achieved with the S reduction in the III$^l$ representation while both reductions in the 
I$^r$ representation required an intermediate number of parameters.

In a study of SO$_2$ isotopologs containing one and two $^{18}$O, asymmetric top molecules 
somewhat close to the prolate limit, Margul{\`e}s et al. \cite{18O-SO2_rot_AuS-red_2020} 
reemphasized Watson's recommendation to favor the S reduction over the A reduction because of 
the smaller correlation coefficients \cite{watson_distortion_1977}. The authors added that 
the S reduction may be a better choice even in cases in which more spectroscopic parameters 
are needed because of the more favorable condition numbers \cite{18O-SO2_rot_AuS-red_2020}. 
The condition number is the ratio of the largest singular value of the Jacobian matrix over 
the smallest and indicates the degree of ill-conditioning; a large value may indicate that the 
determined spectroscopic parameters are not reliable. Among the condition numbers of the fits 
of the main isotopolog, the value is very large, almost 15000, only in the case of the III$^l$ 
representation and the A reduction; the values are much smaller and fairly similar, 150, 91, 
and 92, for the combinations III$^l$ \& S, I$^r$ \& A, and I$^r$ \& S, respectively. Therefore, 
our preference of I$^r$ \& A is supported from the viewpoint of the condition number.

Substitution of one $^{12}$C with $^{13}$C or of one $^{16}$O with $^{18}$O reduces 
the rotational parameters by modest amount, except for $B$ of the $c$-C$_2$H$_4$$^{18}$O 
isotopolog. Substitution of an atom on the symmetry axis does not affect the equilibrium 
rotational parameter in the Born-Oppenheimer approximation. The corresponding ground state 
rotational parameter of a heavy-atom substitution is frequently slightly larger because of 
effects of anharmonicity, as is observed in the case of oxirane. The quartic distortion 
parameters of both minor isotopic species are quite close to those of the main species, 
those of the $^{13}$C species more so than those of the $^{18}$O, which is commensurate 
with the changes in rotational parameters. The situation is more complex among the 
higher order parameters, in part possibly because fewer parameters were varied as a 
consequence of the smaller data sets extended to lower quantum numbers. Some parameters 
of $c$-$^{13}$CCH$_4$O display large changes compared to the main isotopolog, possibly 
as a consequence of the different symmetry. The $\Phi_J$ value of the main species 
appears to be accidentally very small in magnitude compared with almost all of the 
remaining sextic distortion parameters. The $\Phi_J$ value of $c$-$^{13}$CCH$_4$O is 
larger in magnitude and has an opposite sign; it was even necessary to fit the related 
$L_J$ for this isotopolog. On the other hand, $\phi_J$ is much smaller in magnitude than 
that of the main species. The values of some of the octic distortion parameters will 
depend on the values of parameters that were kept fixed in the fit.

We were interested if $c$-$^{13}$CCH$_4$O, as the more promising of the two minor 
isotopologs, could be found in existing astronomical data. The $^{12}$C/$^{13}$C 
ratio is particularly favorable in the Galactic center. Among its sources, 
Sagittarius (Sgr) B2 is among the most prominent sources in the whole Galaxy. 
A search toward Sgr~B2(N1S) and Sgr~B2(N2) in the ReMoCA data, a molecular line 
survey at 3~mm with the Atacama Large Millimeter Array \cite{urea_ReMoCA_2019}, 
was unsuccessful. Assuming a $^{12}$C/$^{13}$C ratio of 20 
\cite{13C-VyCN_2008,EMoCA_2016,RSH_ROH_2016}, all lines of $c$-$^{13}$CCH$_4$O 
were heavily blended with other, stronger lines or did not exceed the 3$\sigma$ 
average noise level of the survey (A. Belloche, private communication, 2021).


\section{Conclusion}
\label{conclusion}

We have obtained greatly improved spectroscopic parameters for the $c$-$^{13}$CCH$_4$O 
and $c$-C$_2$H$_4$$^{18}$O isotopologs which should be sufficiently accurate for all 
radio astronomical observations. The investigations into the performance of various 
combinations of representations and reductions in fitting an extended data set of 
the oxirane main isotopolog revealed once more the poor performance of the A reduction 
in an oblate representation, suggesting that these combinations should be avoided. 
In the more general case of oblate- as well as prolate-type asymmetric rotors 
we conclude that the S reduction should be preferred for fits employing a natural 
choice of representation. Other choices are only recommended if all common combinations 
of representation and reduction were tested.

Concerning $c$-$^{13}$CCH$_4$O in space, it appears as if dedicated searches are required 
to detect this isotopolog in the interstellar medium.

\section*{Note added in proof}

After acceptance of our manuscript, we became aware of an interesting article on 
dimethylsulfoxide \cite{DMSO_2010} that also discusses reductions and representations. 
The findings are in agreement with the findings in publications discussed in our article.

\section*{CRediT authorship contribution statement}

Holger S.P. M\"uller: Conceptualization, Investigation, Methodology, Formal analysis, 
Validation, Data curation, Writing $-$ Original Draft, Writing $-$ review \& editing. 
Jean-Claude Guillemin: Resources, Writing $-$ review \& editing. Frank Lewen: Resources, 
Writing $-$ review \& editing. Stephan Schlemmer: Funding acquisition, Resources, 
Writing $-$ review \& editing.

\section*{Declaration of competing interest}

The authors declare that they have no known competing financial interests or personal 
relationships that could have appeared to influence the work reported in this paper.


\section*{Acknowledgments}

It is our great pleasure to dedicate this work to the memory of J.K.G. Watson for 
his many and invaluable contributions to the theory of high-resolution spectroscopy. 
We will also keep him in mind for his great sense of humor and for his Scottish 
accent. 

We thank Arnaud Belloche for communicating results of his search for $c$-$^{13}$CCH$_4$O 
in astronomical data. 
The work in K{\"o}ln was supported by the Deutsche Forschungsgemeinschaft through the 
collaborative research center SFB~956 (project ID 184018867) project B3 and through 
the Ger{\"a}tezentrum SCHL~341/15-1 (``Cologne Center for Terahertz Spectroscopy''). 
J.-C. G. acknowledges support by the Centre National d'Etudes Spatiales 
(CNES; grant number 4500065585) and by the Programme National Physique et Chimie du 
Milieu Interstellaire (PCMI) of CNRS/INSU with INC/INP co-funded by CEA and CNES. 
Our research benefited from NASA's Astrophysics Data System (ADS).

\appendix
\section*{Appendix A. Supplementary Material}

Supplementary data associated with this article can be found, in the online version, 
at https://doi.org/101016/j.jms. ...



\bibliographystyle{elsarticle-num}
\bibliography{Oxiran}






\end{document}